\documentclass[aip,jvst,reprint]{revtex4-1}
\usepackage{graphicx}
\usepackage{ulem}
\usepackage{amssymb}
\usepackage{amsmath}
\usepackage{wasysym}
\usepackage{relsize}
\begin{document}


\newcommand{\be}{\begin{equation}}
\newcommand{\ee}{\end{equation}}
\newcommand{\bea}{\begin{eqnarray}}
\newcommand{\eea}{\end{eqnarray}}
\newcommand{\Tbar}{{\bar{T}}}
\newcommand{\En}{{\cal E}}
\newcommand{\K}{{\cal K}}
\newcommand{\U}{{\cal U}}
\newcommand{\GC}{{\cal \tt G}}
\newcommand{\Lop}{{\cal L}}
\newcommand{\DB}[1]{\marginpar{\footnotesize DB: #1}}
\newcommand{\q}{\vec{q}}
\newcommand{\kt}{\tilde{k}}
\newcommand{\Lopn}{\tilde{\Lop}}
\newcommand{\noi}{\noindent}
\newcommand{\ovn}{\bar{n}}
\newcommand{\ovx}{\bar{x}}
\newcommand{\ovE}{\bar{E}}
\newcommand{\ovV}{\bar{V}}
\newcommand{\ovU}{\bar{U}}
\newcommand{\ovJ}{\bar{J}}
\newcommand{\calE}{{\cal E}}
\newcommand{\ovphi}{\bar{\phi}}
\newcommand{\zt}{\tilde{z}}
\newcommand{\ttl}{\tilde{\theta}}
\newcommand{\nuv}{\rm v}
\newcommand{\ds}{\Delta s}
\newcommand{\fn}{{\small {\rm  FN}}}
\newcommand{\cc}{{\cal C}}
\newcommand{\cd}{{\cal D}}
\newcommand{\tth}{\tilde{\theta}}
\newcommand{\cb}{{\cal B}}
\newcommand{\cg}{{\cal G}}
\newcommand{\wkb}{\text{WKB}}
\newcommand{\Corr}{\text{Corr}}
\newcommand{\Exact}{\text{Exact}}
\newcommand{\fit}{\text{fit}}
\newcommand{\CCMG}{\text{CCMG}}
\newcommand{\ext}{\text{ext}}
\newcommand\norm[1]{\left\lVert#1\right\rVert}

\title{Higher order curvature corrections to the field emission current density}


\author{Debabrata Biswas}\email{dbiswas@barc.gov.in}
\author{Rajasree Ramachandran}

\affiliation{
Bhabha Atomic Research Centre,
Mumbai 400 085, INDIA}
\affiliation{Homi Bhabha National Institute, Mumbai 400 094, INDIA}

\begin{abstract}
  A simple expression for the Gamow factor is obtained using a
  second order curvature corrected tunneling potential.
  Our results show that it approximates accurately the `exact-WKB' transmission coefficient obtained by
  numerically integrating over the tunneling region to obtain the Gamow factor.
  The average difference in current density using the respective transmission coefficients is
  about $1.5 \%$, across a range of work-functions $\phi \in [3-5.5]$eV, Fermi energy $\calE_F \in [5-10]$eV,
  local electric fields $E_l\in [3-9]$eV
  and radius of curvature $R \geq 5$nm.
  An easy-to-use correction factor $\lambda_P$ is also provided to approximately map the 
  `exact-WKB' current density to the `exact' current density in terms of $\calE_F/\phi$.
  The average error on using $\lambda_P$ is found to be around $3.5\%$ 
  This is a vast improvement over the average error of $15\%$ when $\lambda_P = 1$.
  Finally, an analytical expression for the curvature-corrected current density is
  obtained using the Gamow factor. It is found to compare well with the `exact-WKB'
  current density even at small values of local electric field and radius of curvature.
\end{abstract}

\maketitle

\section{Introduction}
\label{sec:intro}

Field emission refers to the quantum mechanical tunneling of electrons from
the surface of a conductor on application of an external electric field.
\cite{FN,Nordheim,burgess,murphy,jensen2003,forbes,jensen_book,FD2007,DF2008,
  jensen2019,db_rr_2019,db_rk_2019}. It requires local surface fields upwards
of 3V/nm for a measurable current to flow.
Such high fields can be easily achieved
on the surface of sharp tips due to the convergence of field lines resulting in the
enhancement of the local field $E_l$ over the asymptotic or macroscopic applied field, $E_0$
by a factor $\gamma = E_l/E_0$. 
In modern-day usage, $E_l$ embodies the essence of curved emitters and the
field enhancement factor $\gamma$ is the focus of much research
\cite{edgcombe2002,forbes2003,db_fef,db_rudra}. 

In applying the  field emission formalism\cite{FN,murphy,FD2007,jensen_book} to
curved emitters, it is implicitly assumed that
the local field $E_l$ at any point on the surface remains constant along the outward normal
till the end of the classically forbidden region. For $\phi = 4.5$eV and $E_l > 4$V/nm, this is  
typically less that 1.5nm. The corresponding potential
energy is thus expressed as $V_{\ext} = -qE_l s$ where
$s$ is the normal distance from the point on the surface of the emitter.

For a curved emitting tip with apex radius of curvature $R_a < 100$nm, the electric
field can no longer be assumed to be constant in the tunneling region and does
in fact fall off sharply resulting in a reduced tunneling current \cite{db_rr_2019}. The current
density formula must therefore reflect this dependence on $R_a$ accurately
for it to be applied seamlessly in practical applications.

The first correction to the electrostatic potential at the apex of an axially
symmetric emitter was provided in 2015. It was established under general considerations
that along the symmetry axis \cite{KX},

\be
V_{\ext}(s) = -qE_l s \left[1 - \frac{s}{R_a} + \mathcal{O} (\frac{s}{R_a})^2 \right]
\label{eq:kx}
\ee

\noi
where $q$ is the electronic charge.

Using analytically solvable models such as
the hemiellipsoid and hyperboloid diodes, it was subsequently\cite{pop_ext} shown in 2018
that at points close to the apex, the potential has the form

\be
V_{\ext}(s) \approx -qE_l s \left[1 - c_1 \frac{s}{R_2} + c_2 \frac{4}{3}
  \left(\frac{s}{R_2}\right)^2  + \mathcal{O} (s^3) \right] \label{eq:4b3a}
\ee

\noi
where $R_2$ is the second principal radius of curvature at any point
on the surface of the emitter while at the apex, $R_2 = R_a$.
The quantities $c_1$ and $c_2$ depend on the point on the
emitter surface. For points close
to the apex $c_1 \approx 1 \approx c_2$.

Recently\cite{nlcm_pot_ext}, Eq.~(\ref{eq:4b3a}) has been put on a firmer footing using the nonlinear
line charge model\cite{jap2016} for generic smooth emitters. The result has been verified
numerically for various shapes, anode distances and even in the presence of other
emitters. It was found that within about 2 nanometers from the surface of an emitter,
the potential due to the applied external electric field (as determined using COMSOL)
is well approximated by

\be
V_{\ext}(s) = -qE_l s \left[1 -  \frac{s}{R_2}  + \frac{4}{3}
  \left(\frac{s}{R_2}\right)^2  \right]. \label{eq:4b3}
\ee

\noi
At $R_a = 5$nm and $E_l = 3$V/nm, the tunneling region is well within 2nm, while at higher fields,
the width decreases sharply (see for instance Fig.~7 of Ref.~[\onlinecite{nlcm_pot_ext}]).

The potential energy barrier seen by an electron can thus be expressed as

\be
V_T(s) = \phi + V_{\ext}(s) -  \frac{B}{s(1 + \frac{s}{2R_2})} \label{eq:tunnelpot}
\ee

\noi
where the curvature corrected image potential under a locally spherical approximation
has been used with $B = q^2/(16\pi\epsilon_0)$. At $T = 0$,
the free electrons have energy $-\calE_F < \calE \leq 0$. Hereafter, we shall drop
the subscript in $R_2$ and simply using the notation $R$.

A first step towards a curvature-corrected current density
applicable at the apex of an axially symmetric emitter was taken in 2015\cite{KX}. Using
Eq.~(\ref{eq:kx}) for $V_{\ext}$,  an expansion of the
Gamow factor $G$ yields

\be
G = g \int_{s_1}^{s_2} \sqrt{V_T(s) - \calE}~ds = G_0 + x G_1 + \mathcal{O}(x^2)
\ee

\noi
where $ g = \sqrt{8m}/\hbar \simeq 10.246 \text{(eV)}^{-1/2} \text{(nm)}^{-1}$
while $s_1$,$s_2$ are the zeroes of the integrand at $x = (\phi - \calE)/(qE_l R) = 0$.
Since the smallness parameter $x$  tends to zero for large $R$,
$G_0$ represents the Gamow factor in the planar 
limit $R \rightarrow \infty$ or  $x \rightarrow 0$, and
$G_1 = (\partial G/\partial x)_{|x=0}$
is the first curvature dependent correction.
This leads in turn to a curvature corrected current
density applicable at the emitter apex for small values of $x$.

\begin{table}[htb]
 \begin{center}
   \caption{Specification of symbols}
   \vskip 0.25 in
    \label{tab:symbols}
    \begin{tabular}{|l|l|}
      \hline
      $~~~\phi~~~$ & Work function \\ \hline
      $~~~q~~~$ & Magnitude of electron charge \\ \hline
      $~~~\calE_F~~~$ & Fermi energy \\ \hline
      $~~~R_a~~~$ & Apex radius of curvature \\ \hline
      $~~~R_2~~~$ & Second principle radius of curvature \\ \hline
      $~~~E_l~~~$ & Local electric field on emitter surface \\ \hline
      $~~~E_0~~~$ & Macroscopic electric field \\ \hline
      $~~\gamma = E_l/E_0 ~$ & Local field enhancement factor  \\ \hline
      $~~~G~~~$ & Gamow factor in WKB approximation\\ \hline
      $~~G_{\Exact}~~$ & $G$ evaluated numerically  \\ \hline
      $~~G^{(3)}~~$ & Approximate expression for $G$  \\ \hline
      $~~J_{\wkb}^{\Exact}~~$ & Current density using $G_{\Exact}$ \\ \hline
      $~~J_{\wkb}^{(3)}~~$ & Current density using $G^{(3)}$ \\ \hline
      $~~J_{\Exact}~~$ & Current density obtained  numerically \\ \hline        
      $~~\lambda_P^{\fit}~~$ & $J_{\wkb}^{(3)}/J_{\Exact} \approx 0.3400 \calE_F/\phi + 0.3614$ \\ \hline
      $~~J_{\Corr}^{(3)}~~$ & $J_{\wkb}^{(3)}/\lambda_P^{fit}$ \\ \hline
        $~~J_{\wkb}^{(0)}~~$ &  Murphy-Good current density \\ \hline
    \end{tabular}
\end{center}
\end{table}

Our first aim here is to use Eq.~(\ref{eq:4b3}) in the tunneling potential
$V_T$ of Eq.~(\ref{eq:tunnelpot}) and determine a reasonably accurate
expression (referred to as $G^{(3)}$ subsequently)
which approximates $G$ across a wide range of commonly encountered emitter paramaters such as
the workfunction $\phi$, the Fermi energy $\calE_F$, the local field $E_l$ and
the radius of curvature $R$. This is achieved in section \ref{sec:gamow}.
In section \ref{sec:numgamow}, the current density obtained using this transmission
coefficient is compared with the exact-WKB current density ($G$ obtained numerically)
as well as the exact current density (transmission coefficient obtained by solving the Schr\"{o}dinger equation
numerically). Finally, we use
the approximate expression $G^{(3)}$ to arrive at a curvature corrected
analytical expression for the current density $J_{\CCMG}$ in section \ref{sec:analytical}.
Our results are summarized in the concluding section. A list of some of the symbols used
are summarized in Table~\ref{tab:symbols}.

\section{An approximate expression for the Gamow factor}
\label{sec:gamow}

The Gamow factor can be expressed in terms of suitable normalized variables\cite{KX}
$x,y, \text{and}~ \xi$. With $\varphi = \phi - \calE$, they
are expressed as $x = \varphi/(qE_lR)$,
$y = 2\sqrt{B E_l}/\varphi$ and  $\xi = (qE_l/\varphi)z$. Using
Eq.~(\ref{eq:4b3}) in the the tunneling potential of Eq.~(\ref{eq:tunnelpot}), the Gamow factor
can thus be written as 

\be
G(x,y)  =  \frac{2}{3}g\frac{\varphi^{3/2}}{qE_l} \Xi(x,y)
\ee

\noi
where

\be
\begin{split}
\Xi(x,y)  = & \frac{3}{2} \int_{\xi_1}^{\xi_2} d\xi~
\left( 1 - \xi + x \xi^2 - \frac{4}{3}x^2 \xi^3 - \right. \\
& ~ \left. \frac{y^2/4}{\xi + x \xi^2/2} \right)^{1/2}
\end{split}
\ee

\noi
where $\xi_1$ and $\xi_2$ are the real roots of 

\be
1 - \xi + x \xi^2 - \frac{4}{3}x^2 \xi^3 - \frac{y^2/4}{\xi + x \xi^2/2}
\ee

\noi
in the region $\xi > 0$. Note that $x = 0$ gives us the planar result
$\Xi(0,y) = v(y) \approx 1 - y^2 + (y^2/3)\ln(y)$.

For curved emitters (i.e. $x \neq 0$), a possible way forward is to use a Taylor expansion
of $\Xi(x,y)$ 

\be
\Xi(x,y) = \Xi(0,y) + \sum_{k=1}^{k=N} \frac{x^k}{k} \left(\frac{\partial^k \Xi(x,y)}{\partial x^k}\right)_{x=0} + \mathcal{O}(x^4)
\ee

\noi
and hope to achieve convergence in the desired domain of $x$ and $y$,
for instance by restricting to $N = 3$. Since expressions for
each of the partial derivatives have to be evaluated numerically, it is
less cumbersome to evaluate $\Xi(x,y) - \Xi(x,0)$ directly instead of a
Taylor expansion. It is this approach
that we shall adopt here.

\begin{figure}[thb]
\vskip -0.5 cm
\hspace*{-0.75cm}\includegraphics[width=0.62\textwidth]{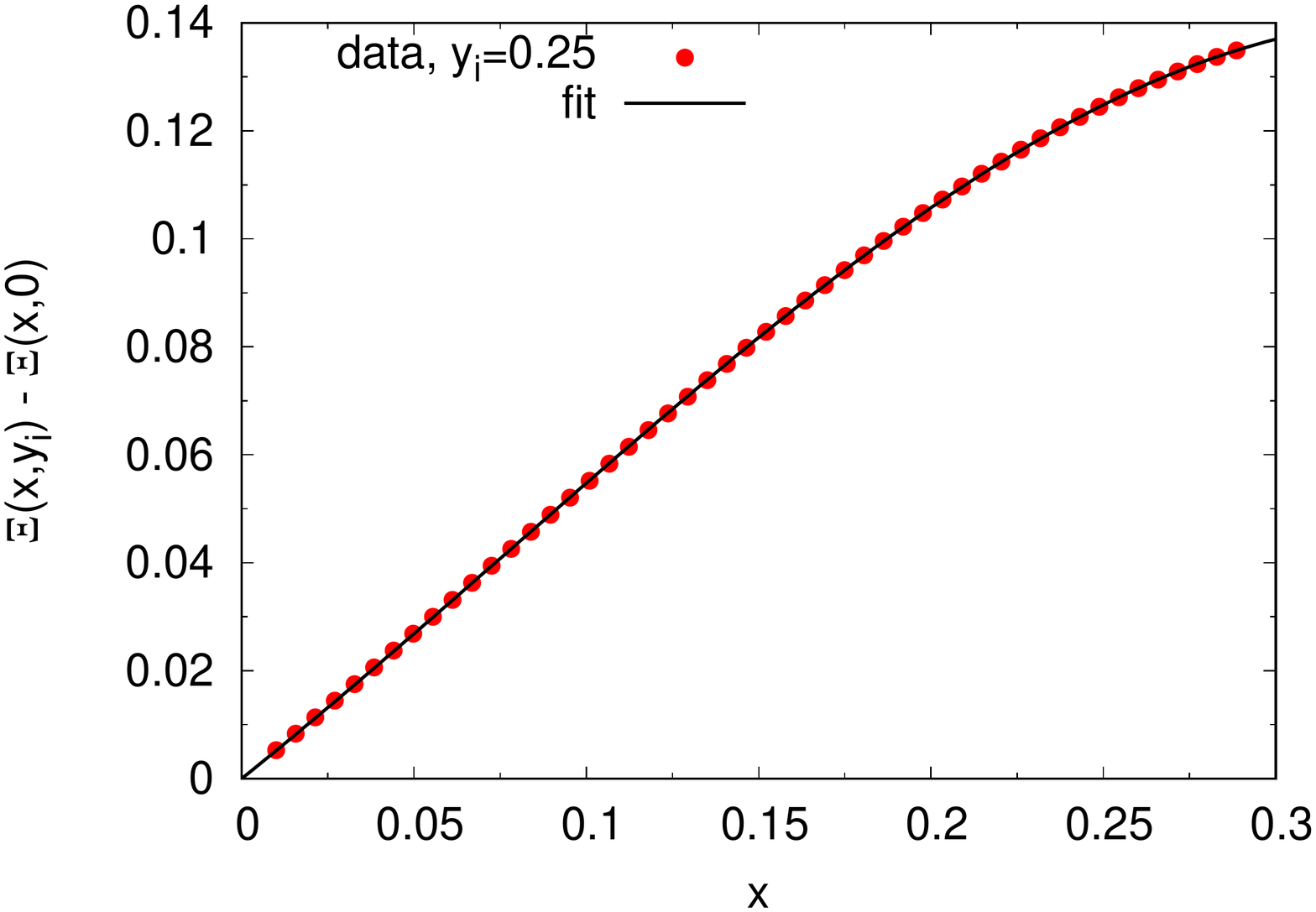}
\vskip -0.75 cm
\caption{Variation of $\Xi(x,y_i) - \Xi(x,0)$ for $y_i = 0.25$ and the fitted
  function.
}
\label{fig:fitX}
\end{figure}

We thus choose a large number of values of $y$ in the interval $(0,1)$. For each
of these $y_i$, we evaluate $\Xi(x,y_i) - \Xi(0,y_i)$ and fit a function

\be
\Xi(x,y_i) - \Xi(0,y_i) \approx x a_{1}^{(i)} + x^2 a_{2}^{(i)} + x^3 a_3^{(i)}
\ee

\noi
and determine the coefficients $a_{k}^{(i)}, k = 1,2,3$. This is then repeated for
all values of $y$ chosen. A typical fit ($y_i = 0.25$) is shown in
Fig.~\ref{fig:fitX}.

\begin{figure}[thb]
\vskip -0.5 cm
\hspace*{-0.75cm}\includegraphics[width=0.62\textwidth]{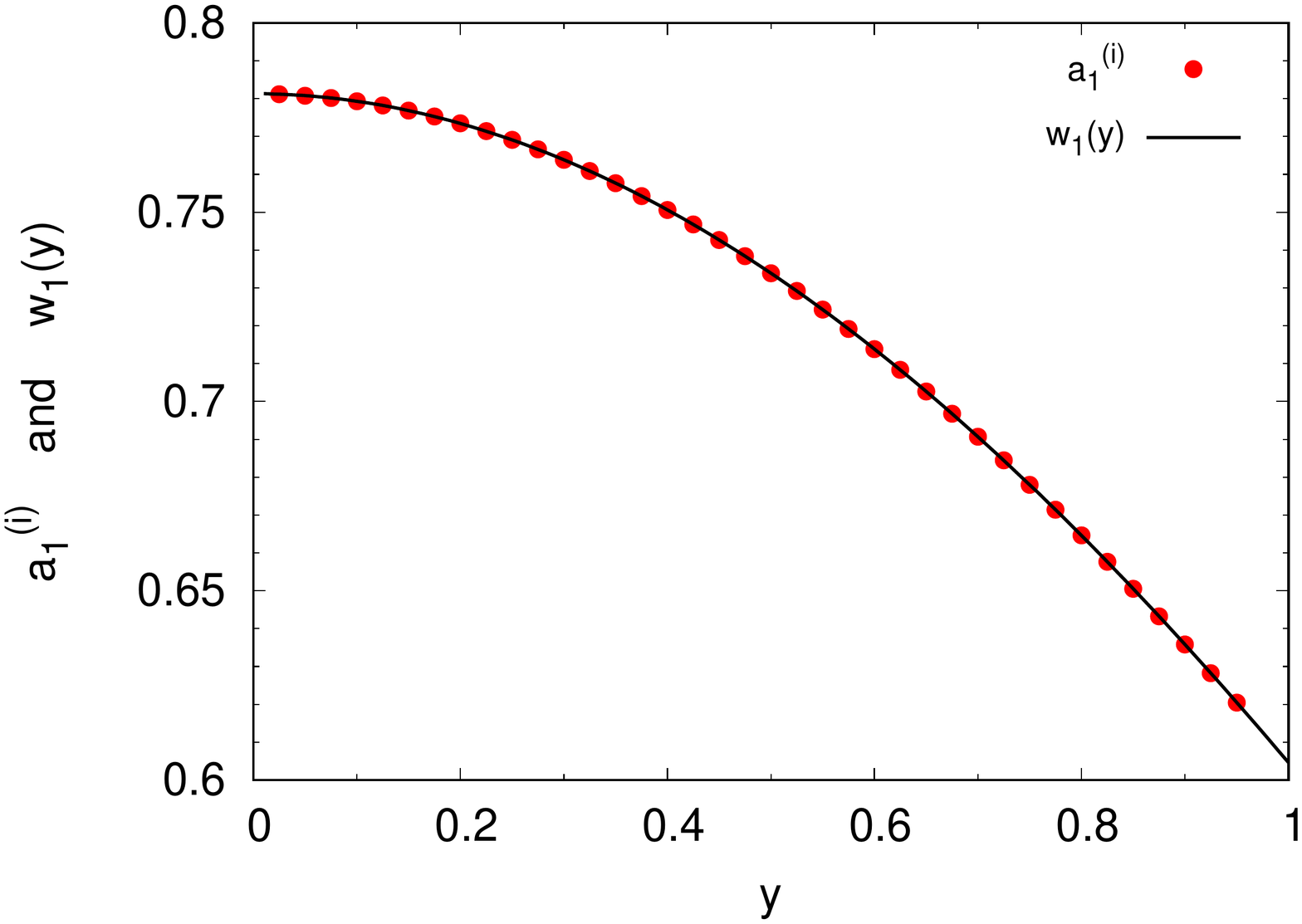}
\vskip -0.75 cm
\caption{Variation of the coefficients $\{a_1^{(i)}\}$ and the fitted
  function $w_1(y)$.
}
\label{fig:w1}
\end{figure}

\begin{figure}[thb]
\vskip -0.5 cm
\hspace*{-0.75cm}\includegraphics[width=0.62\textwidth]{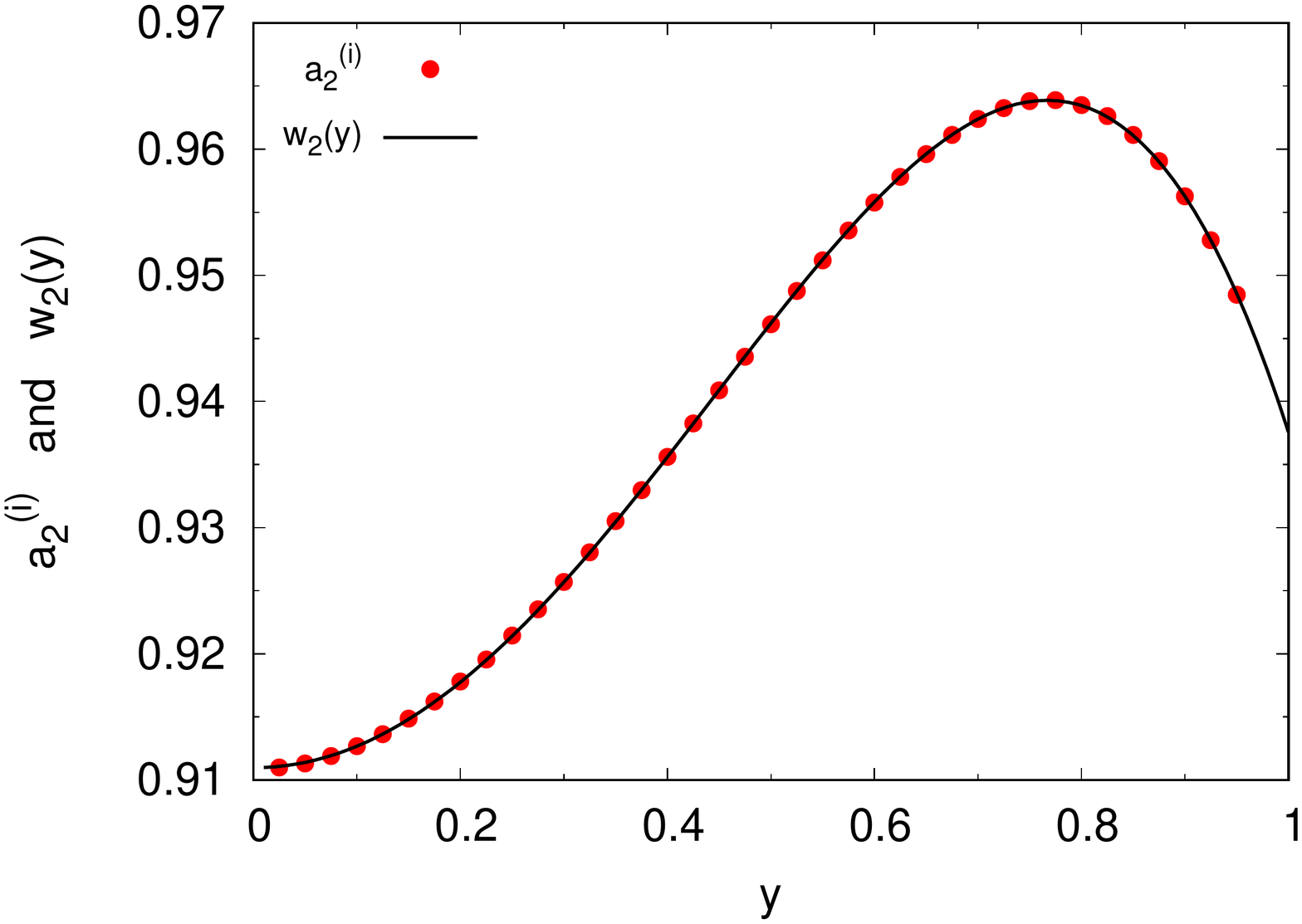}
\vskip -0.75 cm
\caption{Variation of the coefficients $\{a_2^{(i)}\}$ and the fitted
  function $w_2(y)$.
}
\label{fig:w2}
\end{figure}

\begin{figure}[thb]
\vskip -0.5 cm
\hspace*{-0.75cm}\includegraphics[width=0.62\textwidth]{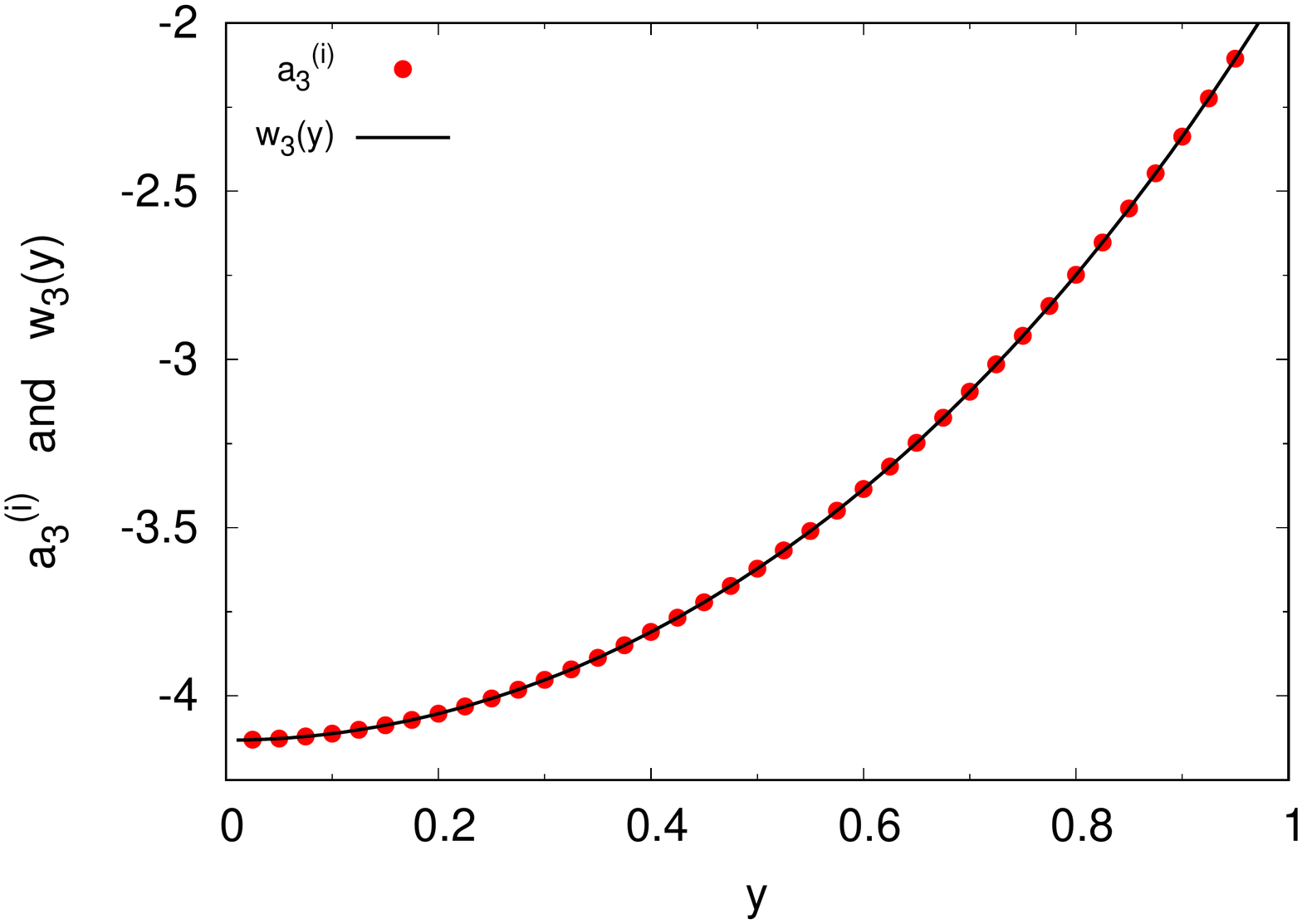}
\vskip -0.75 cm
\caption{ Variation of the coefficients $\{a_3^{(i)}\}$ and the fitted
  function $w_3(y)$.
  }
\label{fig:w3}
\end{figure}

Each of three sets of coefficients $\{a_1^{(i)}\}$, $\{a_2^{(i)}\}$ and $\{a_3^{(i)}\}$
can be used to express their variation with $y$ using a fitting function
in line with the form used for $v(y)$. We choose the functions
$w_k = c_0^{(k)} + c_1^{(k)} y^2 + c_2^{(k)} y^4 + c_3^{(k)} y^2 \ln(y^2)$
and determine the coefficients $c_j^{(k)}, j = 0,1,2,3$
by fitting to $\{a_k^{(i)}\}$. The data $\{a_k^{(i)}\}$ and the fitted functions
$w_k$ are shown in Figs.~\ref{fig:w1} - \ref{fig:w3}.

The fits in each of the three cases is excellent. Note that the quartic term
was required to get a better fit over the entire range of $y$ values.
Also, note that $\{a_1^{i}\}$ are close in value to the function $w(y)$ of
Ref.~[\onlinecite{KX}] but not identical. This is due to the fact that
$\Xi(x,y_i)$ is not restricted to very small values of $x$ while evaluating
the coefficients $\{a_1^{i}\}$. For $x \in [0,0.02]$, the value of $a_1$
for $y=0.025$ is about $0.805 \approx 4/5$ as in Ref.~[\onlinecite{KX}].

The fitting coefficients $c_j^{(k)}, j = 0,1,2,3$ can be suitably approximated so that
as

\bea
w_1(y) & = &\frac{10}{13} - \frac{2}{11}y^2 + \frac{1}{80}y^4 + \frac{1}{400}y^2 \ln(y^2)  \nonumber  \\
w_2(y) & =  & \frac{10}{11} + \frac{2}{11}y^2 - \frac{1}{6}y^4 + \frac{1}{400}y^2 \ln(y^2) \label{eq:ws} \\ 
w_3(y) & =  & -\frac{41}{10} + \frac{39}{20}y^2 + \frac{1}{3}y^4 - \frac{1}{300}y^2 \ln(y^2). 
\eea

\noi
Together with $v(y)$,

\be
G^{(3)}(x,y) = \frac{2}{3}g\frac{\varphi^{3/2}}{qE_l} \left[v(y) + x w_1(y) +
  x^2 w_2(y) + x^3 w_3(y) \right]  \label{eq:gamowfin}
\ee

\noi
and the transmission coefficient $T(\calE)$ can be approximated as

\be
T(\calE) \simeq  e^{-G^{(3)}(x,y)}.   \label{eq:te}
\ee

\noi
Note that the exact tranmission coefficient evaluated by solving the Schr\"{o}dinger equation
numerically differs from Eq.~(\ref{eq:te})
by a factor $P$ which depends on $x$ and $y$.

\section{Comparison with exact results}
\label{sec:numgamow}

Rather than directly comparing the transmission coefficients evaluated
using Eq.~(\ref{eq:ws})-(\ref{eq:te}) with an exact numerical scheme such
as the transfer matrix method \cite{DB_VK}, we shall compare instead
the current densities evaluated using these methods.

The current density within the free electron model can be evaluated
to determine the accuracy of the results. At zero temperature,

\be
  J = \frac{2mq}{(2\pi)^2 \hbar^3} \int_0^{\calE_F} T({\calE'})~ {\calE'} d{\calE'}   \label{eq:Jbasic}
\ee

\noi
where $T({\cal E})$ is the transmission coefficient at electron energy ${\cal E}$ measured with
respect to the top of the conduction band, while $\calE_F$ is the Fermi energy and
$m$ the mass of the electron.

The efficacy of Eq.~(\ref{eq:ws})-(\ref{eq:te}) for evaluating the transmission coefficient $T(\calE)$
can be judged in several ways. In the first instance $J$ can be evaluated using Eq.~(\ref{eq:Jbasic})
and Eq.~(\ref{eq:ws})-(\ref{eq:te}). We shall refer to this as $J_{\wkb}^{(3)}$ to denote that
$T(\calE)$ is evaluated using WKB and incorporates terms upto $x^3$. The current density
can also be evaluated using Eq.~(\ref{eq:Jbasic}) with $T(\calE) = e^{-G_{\Exact}(x,y)}$
where $G_{\Exact}(x,y)$ is determined by numerically integrating
$\Xi(x,y)$ from $\xi_1$ to $\xi_2$. We shall refer to this as $J_{\wkb}^{\Exact}$.
Finally, $J$ can also be evaluated by numerically determining the transmission coefficient
$T(E)$ exactly using the transfer matrix or solving the
Schr\"{o}dinger equation numerically\cite{DB_VK}, and using it in Eq.~(\ref{eq:Jbasic}).
We shall refer to this last evaluation of the current density as $J_{\Exact}$.
 
Each of the three current densities, $J_{\wkb}^{(3)}$, $J_{\wkb}^{\Exact}$ and $J_{\Exact}$ have been
evaluated at 10,000 points respectively corresponding to 10 points each for $\calE_F \in [5,10]$eV,
$\phi \in [3,5.5]$eV, $E_l \in [3,10]$V/nm and $R \in [5,40]$nm. We choose to display these
by plotting the errors as a function of $E_l R/(\calE_F + \phi)$.

\begin{figure}[thb]
\vskip -0.75 cm
\hspace*{-0.75cm}\includegraphics[width=0.62\textwidth]{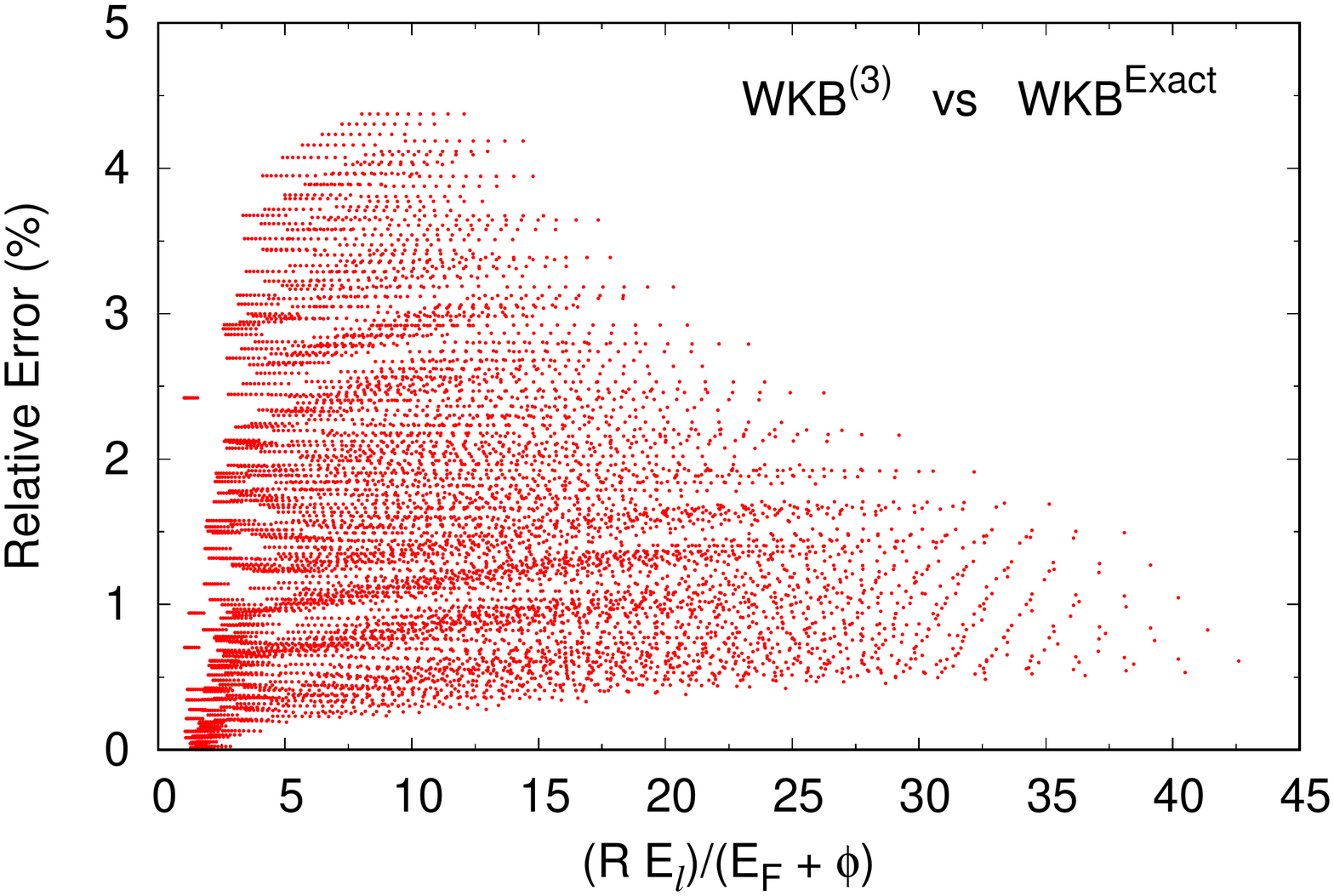}
\vskip -0.75 cm
\caption{A scatter plot of the relative error $100\times|J_{\wkb}^{(3)} - J_{\wkb}^{\Exact}|/J_{\wkb}^{\Exact}$.
  }
\label{fig:wkb3wkbE}
\end{figure}

Fig.~\ref{fig:wkb3wkbE} shows the scatter plot of relative error
$100\times |J_{\wkb}^{(3)} - J_{\wkb}^{\Exact}|/J_{\wkb}^{\Exact}$. Most points have
an error less than $2\%$ while the average error is $1.56\%$.
The maximum error is below $5\%$. 

\begin{figure}[thb]
\vskip -1.0 cm
\hspace*{-0.75cm}\includegraphics[width=0.62\textwidth]{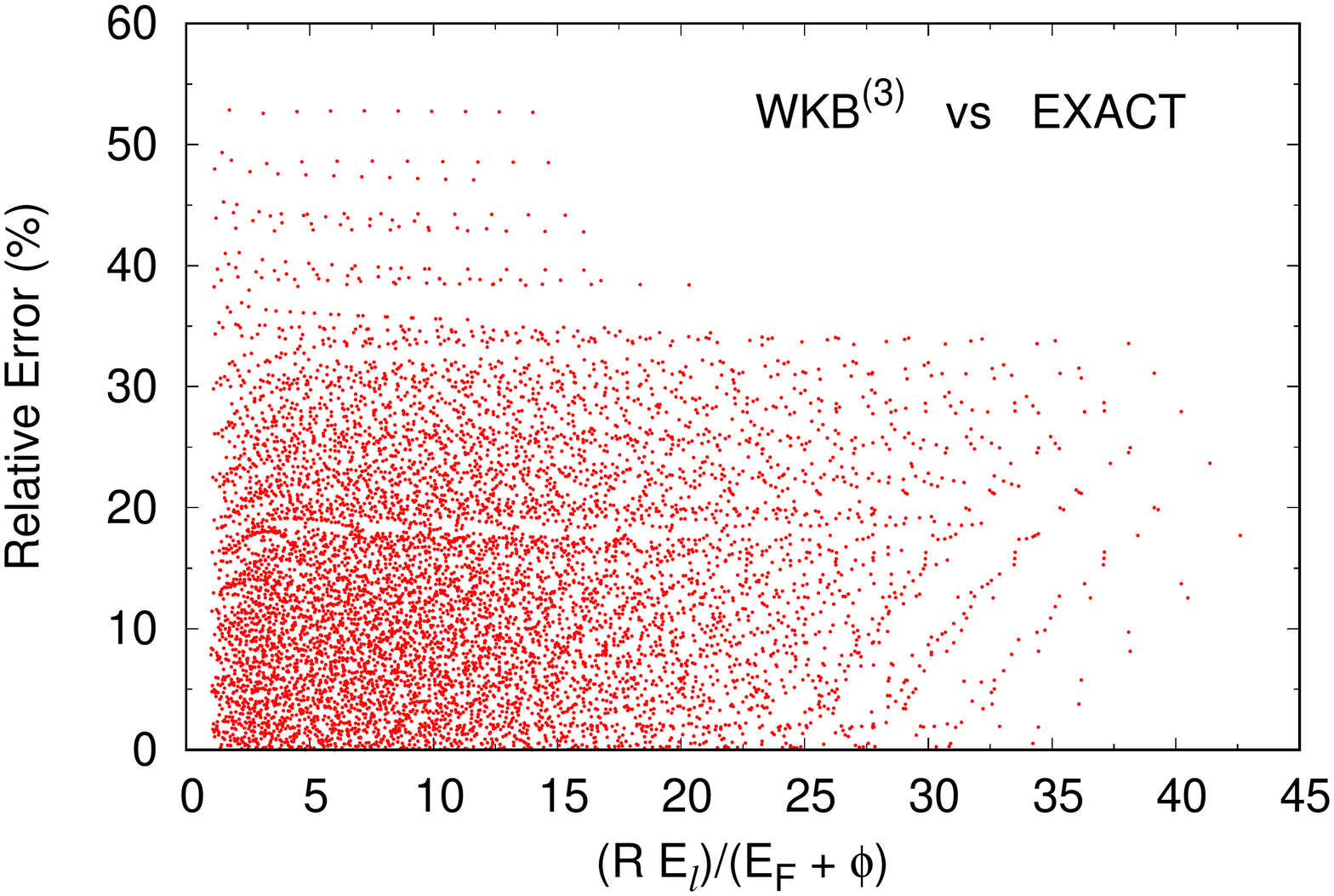}
\vskip -0.75 cm
\caption{A scatter plot of the relative error $100\times|J_{\wkb}^{(3)} - J_{\Exact}|/J_{\Exact}$.
  }
\label{fig:wkb3E}
\end{figure}

Fig.~\ref{fig:wkb3E} shows the scatter plot of relative error
$100\times |J_{\wkb}^{(3)} - J_{\Exact}|/J_{\Exact}$.  The errors can be large as seen from the
scatter plot with the maximum error at around $55\%$ while the average error is
$14.66\%$. The difference is due to the factor $P$ which relates the exact and
WKB transmission coefficients.

In order to reduce the large difference between $J_{\wkb}^{(3)}$ and $J_{\Exact}$, we shall study the
ratio $\lambda_P = J_{\wkb}^{(3)}/J_{\Exact}$. Clearly $\lambda_P$ depends on
$\calE_F, \phi, E_l,~\text{and}~R$. An expansion of $\lambda_P$ in terms of these
variables along the lines of Ref.~[\onlinecite{Mayer2011}] can in principle be carried out.
Our interest here to find a simple expression for $\lambda_P$ that is easy to use and
reduces the average error substantially.

\begin{figure}[thb]
\vskip -1.0 cm
\hspace*{-0.75cm}\includegraphics[width=0.62\textwidth]{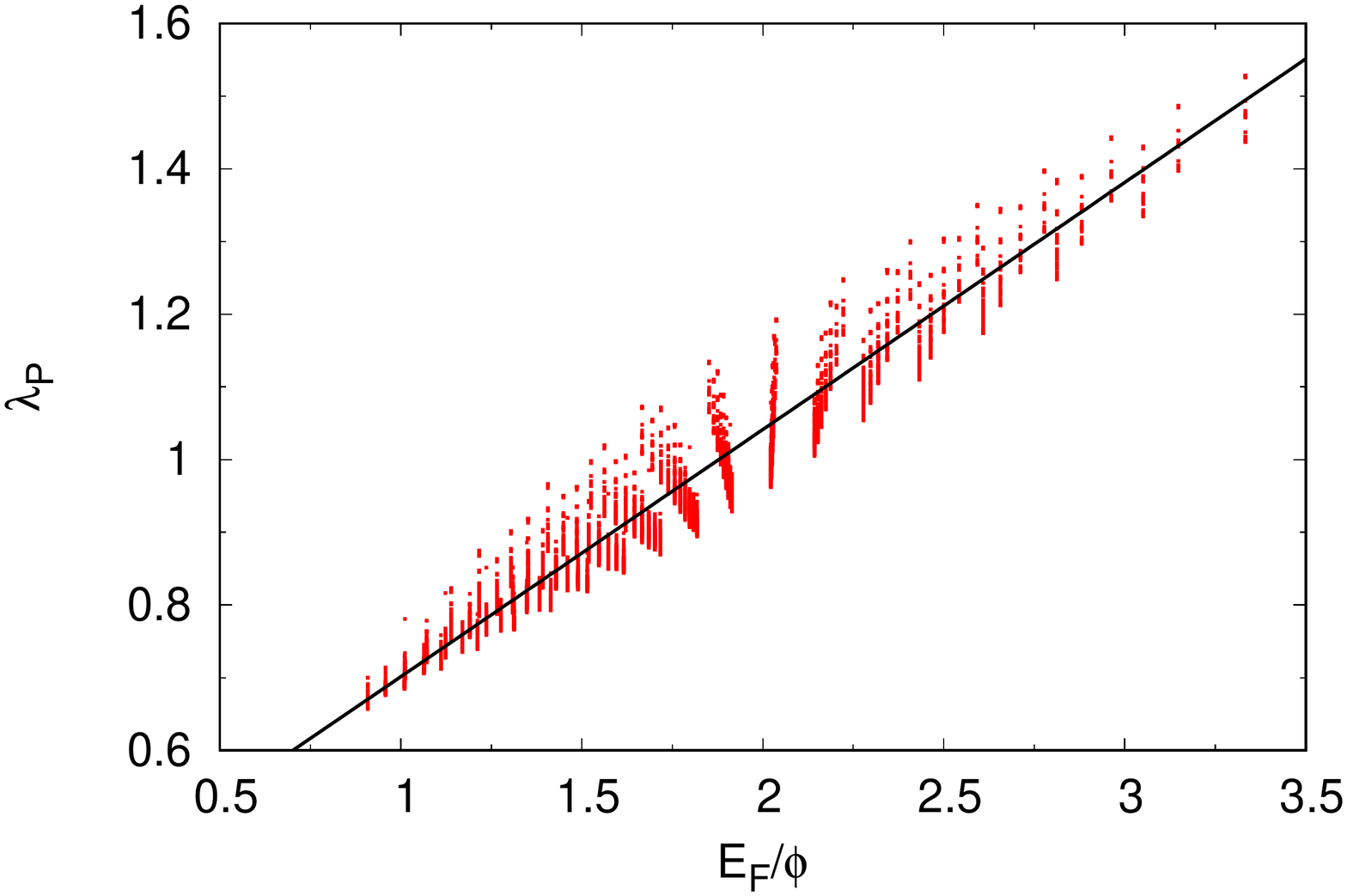}
\vskip -0.75 cm
\caption{The factor $\lambda_P = J_{\wkb}^{(3)}/J_{\Exact}$ plotted
  against $\calE_F/\phi$. Also shown is the best fitting straight line.
  }
\label{fig:prefact}
\end{figure}

Fig.~\ref{fig:prefact} shows  $\lambda_P = J_{\wkb}^{(3)}/J_{\Exact}$
plotted against $\calE_F/\phi$. Clearly, the points seems to display a linear
increase with $\calE_F/\phi$. The best fitting straight line $\lambda_P^{\fit} = a_0 \calE_F/\phi + b_0$
is also shown in figure with $a_0 \approx 0.3400$ and $b_0 \approx 0.3614$.

\begin{figure}[thb]
\vskip -1.0 cm
\hspace*{-0.75cm}\includegraphics[width=0.62\textwidth]{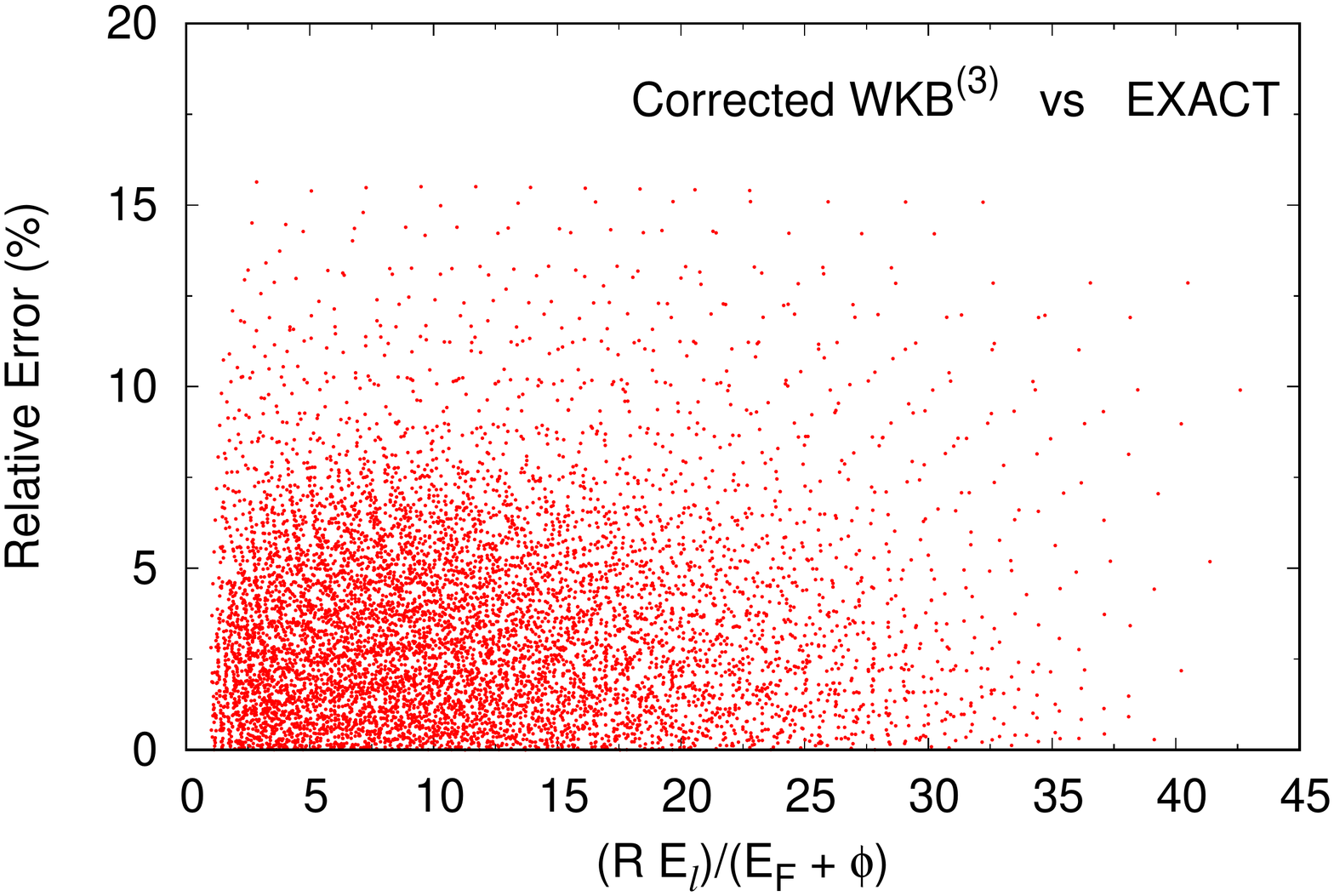}
\vskip -1.0 cm
\caption{A scatter plot of the relative error $100\times |J_{\Corr}^{(3)} - J_{\Exact}|/J_{\Exact}$.
  }
\label{fig:corr}
\end{figure}

The corrected current density $J_{\Corr}^{(3)} = J_{\wkb}^{(3)}/\lambda_P^{\fit} = 
J_{\wkb}^{(3)}/(a_0 \calE_F/\phi + b_0)$
does not eliminate the error completely but is expected to reduce it
substantially. Fig.~\ref{fig:corr} is a scatter plot of the relative error
$100\times |J_{\Corr}^{(3)} - J_{\Exact}|/J_{\Exact}$. The average error is $3.58\%$
and the maximum error has also reduced substantially.

Clearly the curvature corrected WKB current density $J_{\wkb}^{(3)}$
is accurate over a wide range of parameters. It
can be used along with $\lambda_P^{\fit}$ to get a reasonably accurate current
density $J_{\Corr}^{(3)}$ which compares well with the exact current density.

\vskip 0.25 in
\section{Corrections to the current density}
\label{sec:analytical}

The curvature-corrected zero-temperature current density can be evaluated using Eq.~(\ref{eq:Jbasic}),
Eq.~(\ref{eq:te}) and Eq.~(\ref{eq:gamowfin}). The prefactor $\lambda_P^{\fit}$ can be
optionally used to get a more accurate expression.

\subsection{An analytical expression}
\label{subsec:ana}

Note that the dominant contribution to the integral in Eq.~(\ref{eq:Jbasic}) is close
to the Fermi level ($\calE' = 0$) since $T(\calE')$ decreases sharply with $\calE'$.
A Taylor expansion of $G$ at $\calE = 0$ is therefore used\cite{murphy}. Thus,

\be
G(\calE) = G(0) + \calE \left(\frac{\partial G}{\partial \calE} \right)_{\calE = 0} + \mathcal{O}(\calE^2). \label{eq:Gexpand}
\ee

\noi
An expression for $(\frac{\partial G}{\partial \calE})_{|_{\calE = 0}}$ can be easily obtained by
noting that $(\frac{\partial x}{\partial \calE})_{|_{\calE = 0}} = -x_F/\phi$ and
$(\frac{\partial y}{\partial \calE})_{|_{\calE = 0}} = y_F/\phi$ where
$x_F = \phi/(q E_l R)$ and $y_F = 2\sqrt{B E_l}/\phi$. Thus

\be
\begin{split}
  \left(\frac{\partial G}{\partial \calE}\right)_{\calE = 0}  = & -\frac{1}{d_F}  \left[ \left\{ v(y_F) - \frac{2}{3} y_F \left( \frac{dv}{dy}\right)_{y = y_F}\right\} + \right. \\
  & \left.  x_F\left\{\frac{5}{3}w_1(y_F) - \frac{2}{3} y_F \left( \frac{dw_1}{dy}\right)_{y = y_F}\right\}  + \right. \\
  & \left. x_F^2\left\{\frac{7}{3}w_2(y_F) - \frac{2}{3} y_F \left( \frac{dw_2}{dy}\right)_{y = y_F}\right\} + \right. \\
  & \left. x_F^3\left\{3w_3(y_F) - \frac{2}{3} y_F \left( \frac{dw_3}{dy}\right)_{y = y_F}\right\} \right]
  \end{split}
\ee

\noi
where $d_F^{-1} = g\phi^{1/2}/(q E_l)$. This can be further simplified and expressed as

\be
\left(\frac{\partial G}{\partial \calE}\right)_{\calE = 0} = -\frac{t(y_F) + x_F\psi_1(y_F) + x_F^2\psi_2(y_F) + x_F^3 \psi_3(y_F)}{d_F}
\ee

\noi
where on using $y_F^2 = f$,

\bea
t & = & 1 + \frac{f}{9} - \frac{1}{18}f \ln(f) \\
\psi_1 & = & \frac{25}{13} - \frac{237}{1100} f - \frac{1}{480}f^2 - \frac{7}{2400} f \ln(f) \\
\psi_2 & = & \frac{70}{33} + \frac{589}{3300} f + \frac{1}{18}f^2 + \frac{1}{400} f \ln(f) \\
\psi_3 & = & -\frac{123}{10} + \frac{2929}{900} f + \frac{1}{9}f^2 - \frac{1}{180} f \ln(f).
\eea

\noi
Thus, finally using Eqns.~(\ref{eq:te})-(\ref{eq:Gexpand}), the current
density can be finally expressed as

\be
J = \frac{2mq}{(2\pi)^2\hbar^3} \frac{d_F^2}{t_c^2} e^{- v_c B_{\small FN} \phi^{3/2}/E_l}
\ee

\noi
where $v_c$ and $t_c$ are the correction factors due to emitter curvature and image potential.
They are

\be
\begin{split}
  v_c(f) = &  \left( 1 - f + \frac{1}{6} f\ln(f) \right) + \\
  & x_F \left(\frac{10}{13} - \frac{2}{11}f + \frac{1}{80}f^2 + \frac{1}{400}f \ln(f) \right) + \\
  & x_F^2 \left(\frac{10}{11} + \frac{2}{11}f - \frac{1}{6}f^2 + \frac{1}{400}f \ln(f) \right) + \\
  & x_F^3 \left(-\frac{41}{10} + \frac{39}{20}f + \frac{1}{3}f^2 - \frac{1}{300}f \ln(f) \right)
\end{split} \label{eq:vc}
\ee

\noi
and

\be
\begin{split}
t_c(f) & =  \left( 1 + \frac{f}{9} - \frac{1}{18}f \ln(f) \right) +  \\
& x_F \left(\frac{25}{13} - \frac{237}{1100} f - \frac{1}{480}f^2 - \frac{7}{2400} f \ln(f) \right) +  \\
& x_F^2 \left( \frac{70}{33} + \frac{589}{3300} f + \frac{1}{18}f^2 + \frac{1}{400} f \ln(f) \right) + \\
& x_F^3 \left( -\frac{123}{10} + \frac{2929}{900} f + \frac{1}{9}f^2 - \frac{1}{180} f \ln(f) \right).
\end{split}  \label{eq:tc}
\ee

The final expression for the
curvature-corrected-Murphy-Good (CCMG) current density in terms of the
conventional Fowler-Nordheim constant $A_\fn$ and $B_\fn$ takes the form

\be
J_{\CCMG} =  \frac{{\small A_{\small FN}}}{{\small \phi}} \frac{E_l^2}{{t_{c}}^2} \exp\left(-v_{c} B_{\small FN} \phi^{3/2}/E_l \right)  \label{eq:Jcorr}
\ee

\noi
where $E_l$ refers to the local electric field on the emitter surface,
$A_\fn~\simeq~1.541434~{\rm \mu A~eV~V}^{-2}$,
$B_\fn~\simeq 6.830890~{\rm eV}^{-3/2}~{\rm V~nm}^{-1}$,
$f  \simeq  1.439965~E_l/\phi^2$ and $x_F = \phi/(qE_l R)$.

Equations (\ref{eq:vc})-(\ref{eq:Jcorr}) provide a direct means of
evaluating the local current density instead of numerically evaluating the integral
in Eq.~(\ref{eq:Jbasic}).

\subsection{Comparison with exact-WKB result}
\label{subsec:num}

It may be noted that the curvature-corrected Murphy-Good
formula of Eq.~(\ref{eq:Jcorr}) is an approximation of the integral
in Eq.~(\ref{eq:Jbasic}) evaluated using Eq.~(\ref{eq:te}), and
is not necessarily accurate.
At smaller values of the local field, electrons close to the Fermi level
are expected to contribute and hence Eq.~(\ref{eq:Gexpand}) is likely to
be adequate. With an increase in  local field strength, electrons further
away from the Fermi level can tunnel through due to the decrease in height
and width of the barrier. The truncation of the series in
Eq.~(\ref{eq:Gexpand}) may thus lead to errors at higher applied fields.
Such a problem in fact exists even in the 
commonly used  Murphy-Good
expression for current density as we shall see. The energy-integration error
is expected to assume significance in $J_{\CCMG}$ for large $R$ while
errors at small $R$ may be due to curvature effects.

In the following, we shall test how well $J_{\CCMG}$  approximates $J_{\wkb}^{\Exact}$
across a range of radius of curvature given that there are two levels of
approximation involved in going  from $J_{\wkb}^{\Exact}$ to
$J_{\CCMG}$ given by Eq.~(\ref{eq:Jcorr}). The first of these is
the use of $G^{(3)}$ instead of $G_{\Exact}$ while the second involves the
energy integration mentioned above. It is the combined effect of the
two that will manifest as the relative error.

\begin{figure}[hbt]
\vskip -1.0 cm
\hspace*{-0.75cm}\includegraphics[width=0.6\textwidth]{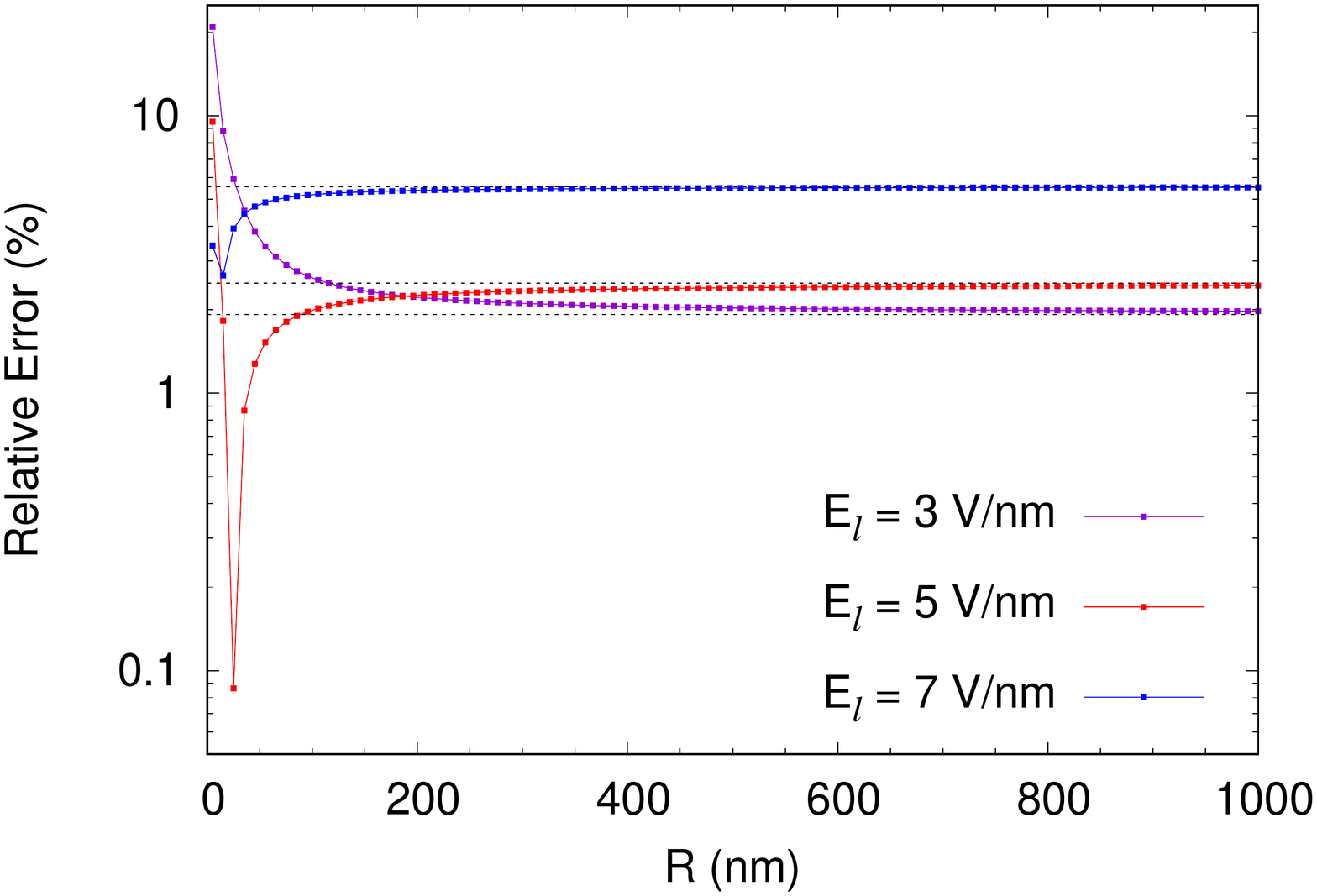}
\vskip -0.75 cm
\caption{Variation in relative error $100 \times |J_{\CCMG} - J_{\wkb}^{\Exact}|/J_{\wkb}^{\Exact}$ with
radius $R$. Here $\phi = 4.5$eV and $\calE_F = 8.5$eV.
  }
\label{fig:analy3}
\end{figure}

\begin{figure}[hbt]
\vskip -1.0 cm
\hspace*{-0.75cm}\includegraphics[width=0.6\textwidth]{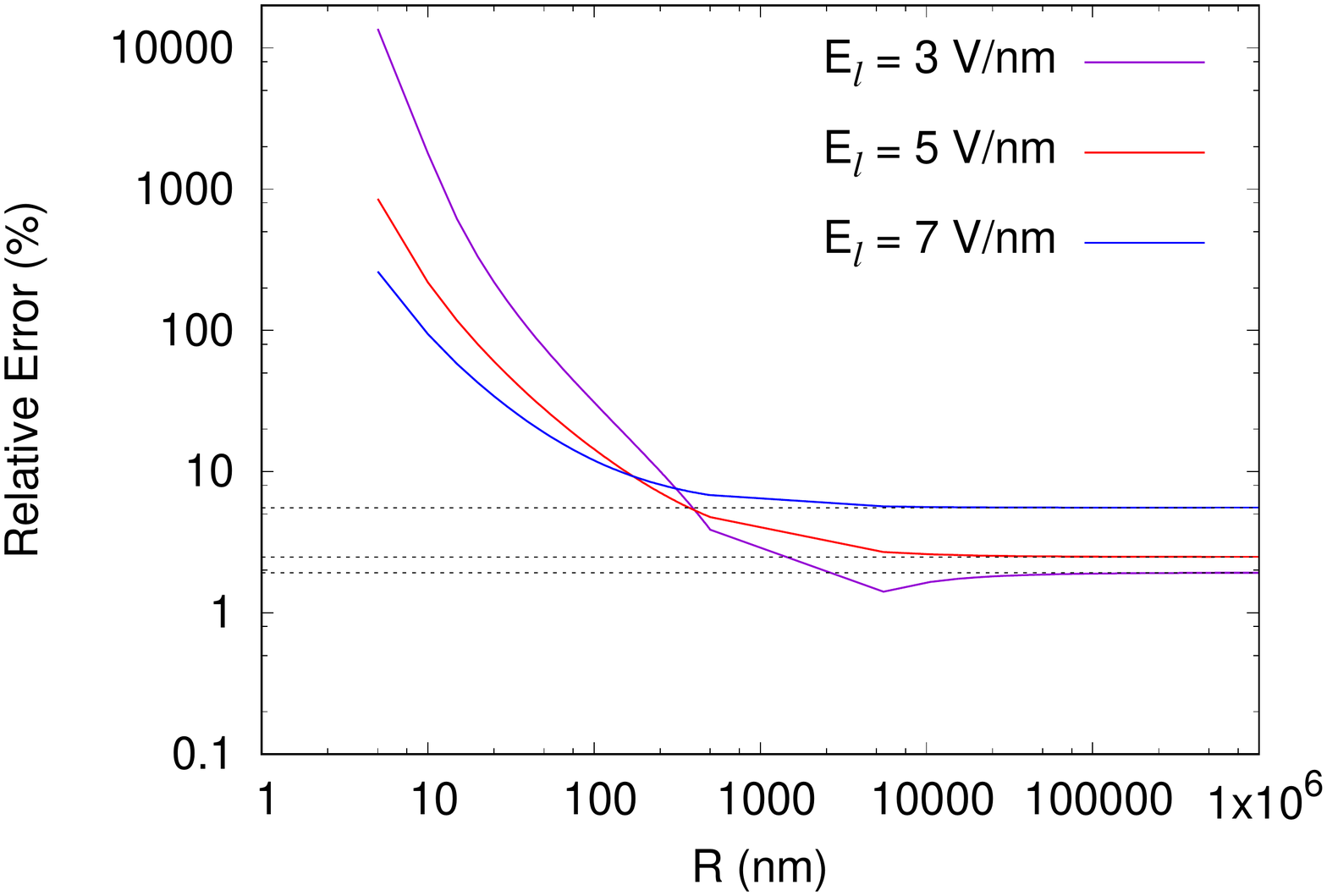}
\vskip -0.75 cm
\caption{Variation in relative error $100 \times |J_{\wkb}^{(0)} - J_{\wkb}^{\Exact}|/J_{\wkb}^{\Exact}$ with
radius $R$. Here $\phi = 4.5$eV and $\calE_F = 8.5$eV.
  }
\label{fig:analy_forbes}
\end{figure}

For simplicity, we consider a fixed value $\phi = 4.5$eV and $\calE_F = 8.5$eV but
three different values of the local field, $E_l$. The radius $R$ is varied and the
relative error $100 \times |J_{\CCMG} -  J_{\wkb}^{\Exact}|/J_{\wkb}^{\Exact}$
is recorded. The results are shown in Fig.~\ref{fig:analy3}. The dashed lines
mark the error in $J_{\CCMG}$ at large $R$ and their values coincide with the error
in planar Murphy-Good result (i.e. using $x_F = 0$ in
Eq.~(\ref{eq:Jcorr})) for different values of the local field.
As expected, the asymptotic error ($R$ large) is small at $E_l = 3$V/nm but increases
with $E_l$. On the other hand, at smaller radius of curvature ($R < 50$nm), the error is small at higher
values of the local field ($x_F$ small) but increases at lower values of
local field. This is again expected since $x_F$ becomes larger as $E_l$ decreases
and $G^{(3)}$ does not approximate $G_{\Exact}$ as well. For $R > 50$nm, the
error due to energy integration seems to dominate and appears to be the limiting
factor.

It is also instructive to see how well the curvature-uncorrected current density
$J_{\wkb}^{(0)}$ fares in dealing with the curvature-dependent tunneling potential.
Fig.~\ref{fig:analy_forbes} shows the relative error
$100 \times |J_{\wkb}^{(0)} - J_{\wkb}^{\Exact}|/J_{\wkb}^{\Exact}$ where $J_{\wkb}^{(0)}$
is evaluated by setting $x_F = 0$ in
Eq.~(\ref{eq:Jcorr}), while $J_{\wkb}^{\Exact}$ is evaluated as before using
the tunneling potential of Eq.~(\ref{eq:tunnelpot}) with the curvature corrections.
The convergence to the strictly planar error limit is slow for each of the three
local fields. At small $R$, the error is very high in each of three cases with the lowest
field having the highest error as expected.

\begin{figure}[hbt]
\vskip -1.0 cm
\hspace*{-0.75cm}\includegraphics[width=0.6\textwidth]{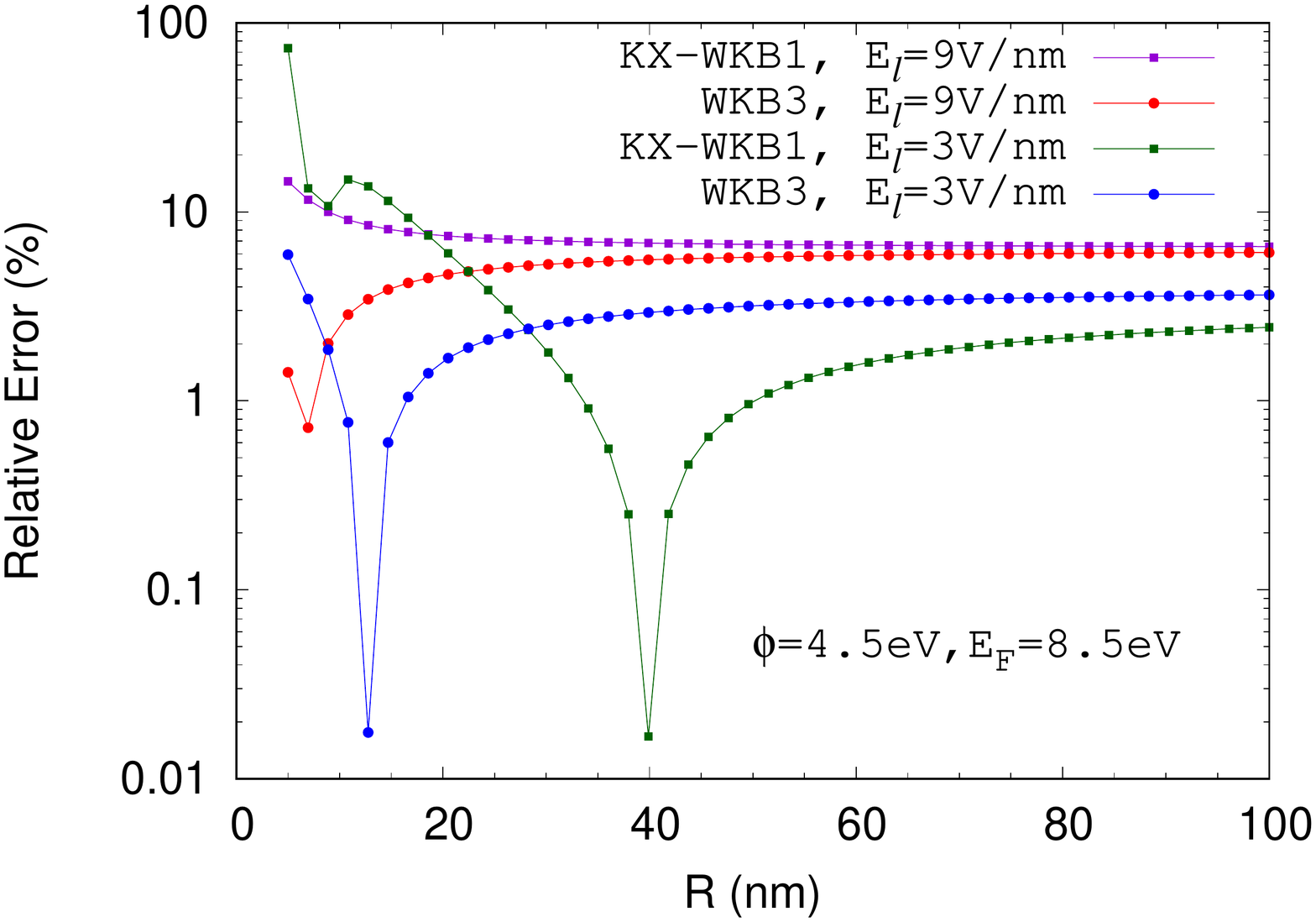}
\vskip -0.75 cm
\caption{Variation in relative error  with
radius $R$. Here $\phi = 5.0$eV and $\calE_F = 8.5$eV.
  }
\label{fig:analy_kyrit}
\end{figure}

The first order curvature-corrected current density
of Kyritsakis and Xanthakis\cite{KX} (denoted by KX-WKB1)
can also be compared to $J_{\wkb}^{\Exact}$ since a Taylor expansion
of the Gamow factor at $x = 0$ using the tunneling potential of
Eq.~(\ref{eq:tunnelpot}) gives identical result. The result is expected to
be good at larger local fields and radius of curvature.
Figure \ref{fig:analy_kyrit} shows a comparison of the relative errors
for KX-WKB1 and $J_{\CCMG}$ (denoted by WKB3) with respect to
$J_{\wkb}^{\Exact}$ at 2 different local fields. For KX-WKB1,
$v_c = (1 - f + \frac{f}{6} \ln f) + x_F (\frac{4}{5} - \frac{7}{40} f
+ \frac{1}{200}\ln f)$ while $t_c = (1 + \frac{f}{9} - \frac{f}{18} \ln f) +
x_F (\frac{4}{3} - \frac{f}{15} - \frac{f \ln f}{1200})$.
Note again that energy-integration errors begin to dominate in both
cases with an increase in radius of curvature while at smaller values of
$R$, this error is somewhat suppressed since the barrier is wider and
Eq.~(\ref{eq:Gexpand}) may be adequate.
It is however clear that for small $R$ and $E_l$, there is a clear
advantage in using $J_{\CCMG}$.

\section{Conclusions}

We have considered higher order curvature corrections to the current density
for the near-universal tunneling potential of Eq.~(\ref{eq:tunnelpot}),
which contains quadratic and cubic curvature-dependent terms in the external potential.

This has been achieved by first finding a suitable expression which accurately
represents the exact Gamow factor for a range of local field, radius of curvature,
Fermi energy, work function and electron energy. This was used to compute
the current density by integrating over the energy states numerically and compared
with the exact-WKB current density. The errors were found to be small
for a range of parameters thereby validating the expression for the
Gamow factor presented in this paper.
We then proceeded to determine an expression for the
curvature-corrected current density $J_{\CCMG}$,
following the standard procedure of expanding the transmission coefficient
at the Fermi level and carrying out the energy integration in Eq.~(\ref{eq:Jbasic}).
The errors relative to the exact-WKB current density was again found
to be reasonably small. In comparison, the standard planar Murphy-Good result
was found to have very large error at smaller radius of curvature and local fields.
For larger radius of curvature ($R > 50$nm), the error in the curvature-corrected
current density $J_{\CCMG}$ appears to be dominated by the errors in the
energy integration.

We also studied the problem of discrepancy between the `exact'
and exact-WKB current densities and discovered a pattern in their ratio when plotted
against $\calE_F/\phi$. This enabled us to suggest a simple prefactor $\lambda_P^{\fit}$ to
correct the exact-WKB current density. The average error over a range of
parameters was found to be about $3.58\%$. Keeping in mind the underlying uncertainties
in modeling the geometric and material properties of emitters and their effect
on the field emission current, errors below $10\%$ are clearly acceptable.

Finally, the curvature-corrected expression for current density $J_{\CCMG}$ can be combined
with the cosine law\cite{db_ultram,physE}  of variation of the local field on the surface of
generic parabolic  emitter tips, for determining an approximate expression for the total
field emission current and distributions of emitted particles with respect to launch angle,
total and normal energy along the lines of Ref. [\onlinecite{parabolic}] where an
emitter with large apex radius of curvature was considered.

\section{Acknowledgements}

The authors acknowledge useful discussions with Dr. Raghwendra Kumar.

\vskip 0.25 in
\noi
{\it Data Availability}: The computational data that supports the findings of this study are available within the article.

\section{References} 



\begin{thebibliography}{99}
\bibitem{samesoln} In a Taylor expansion of the Gamow factor at $x = 0$, the first
  $x$ dependent term is the same as in [\onlinecite{KX}] even for the full
  tunneling potential which includes $\frac{4}{3} qE_l \frac{s^3}{R^2}$.
\end{thebibliography}
\end{document}